# Conditional entropic approach to nonequilibrium complex systems with weak fluctuation correlation

Yuichi Itto [1,2]

[1] Science Division, Center for General Education, Aichi Institute of Technology,
 Aichi 470-0392, Japan

[2] ICP, Universität Stuttgart, 70569 Stuttgart, Germany

**Abstract:**  A conditional entropic approach is discussed for nonequilibrium complex systems with a weak correlation between spatiotemporally fluctuating quantities on a large time scale. The weak correlation is found to constitute the fluctuation distribution that maximizes the entropy associated with the conditional fluctuations. The approach is illustrated in diffusion phenomenon of proteins inside bacteria. A further possible illustration is also presented for membraneless organelles in embryos and beads in cell extracts, which share common natures of fluctuations in their diffusion.

# 1. Introduction

Consider a nonequilibrium complex system divided into a lot of small spatial regions or "blocks", each of which is in a local equilibrium state characterized by dynamics of two different fluctuating quantities (e.g., the local temperature). The system is globally in nonequilibrium-stationary-state-like situation. Let $x$ and $y$ be random variables expressing two such quantities. The time scale of their dynamics is much larger than that of a typical local dynamics (e.g., the one of a random walker): the dynamics in the former, i.e., $x$ and $y$, are slow, whereas the one in the latter is fast. The joint probability distribution is given here by $g(x,y) = g(x|y) f(y)$, where $g(x|y)$ is the conditional distribution describing the probability of $x$, given a value of $y$, and $f(y)$ is the marginal distribution describing the probability of $y$. The conditional entropy associated with $g(x|y)$ at a given value of $y$, say $y_0$, is then as follows:

$$S[g] = -\int dx\, g(x|y_0) \ln g(x|y_0), \tag{1}$$

which is of the form of the Shannon entropy [1].

In recent years, various discussions have been developed about using the maximum entropy principle [2] for treating nonequilibrium complex systems organized hierarchically by different dynamics on largely separated time scales, e.g., as in Refs. [3-10]. For the joint distributions describing these systems, there often exists a situation that the distribution to be examined is the marginal one while the conditional distribution is well understood, which is somewhat in contrast to the situation we consider here. As will be seen later, the conditional entropy in Eq. (1) is shown to play a role for determining the conditional fluctuation distribution mentioned above.

An illustrative example of such systems is a Brownian particle moving through a fluid environment that is divided into many small spatial regions like in the above, over which the (inverse) temperature locally fluctuates from one region to another, see e.g., Ref. [4]. The time scale of the variation of the temperature fluctuations is much larger than that of the relaxation of the particle in a region to a local equilibrium state. So, the effective energy in the region is the fast variable, whereas fluctuating inverse temperature is the slow one. (It is noted that the latter is consistent with $x$ or $y$ as will be discussed later, whereas the former is not.) A stationary state of the region is then given by the joint distribution based on the conditional distribution describing a local equilibrium state at a given value of the inverse temperature and the marginal distribution of the temperature fluctuations. (For the joint distribution, the integration over the inverse temperature leads to a "statistics of statistics" known as superstatistics [11], which will be mentioned later in our present context.) With these distributions, the joint entropy concerning both the fast variable, i.e., the effective energy, and the slow variable, i.e., fluctuating inverse temperature, or the marginal entropy relevant to the slow one has been treated as the quantity to be maximized, for example.

The direction of maximizing such entropies seems to be pertinent due to the existence of the time-scale separation. In our present context, the system under consideration is in nonequilibrium-stationary-state-like situation as mentioned earlier and accordingly each local region in the system is in a quasi-equilibrium state to be described by a canonical ensemble, in which fluctuating local temperature attains a certain value: the stationarity is supposed to hold under the large separation of time scales. Therefore, it is natural to maximize the entropy based on such a "quasi-canonical ensemble". From this perspective, the conditional entropy in Eq. (1) is considered to be maximized, here.

It may be worth mentioning that the problem of studying dynamical systems similar to the above ones has been discussed in the field of social sciences [12], in which the maximum entropy principle is crucial. For example, in Ref. [13], a system such as an urban/regional one consists of several local settlements interlinked each other, each of which contains elements, e.g., residents. The elements are self-reproduced in the settlements and are redistributed among them, where the relaxation time of the redistribution process is much faster than that of the self-reproduction one. Then, the conditional maximization of entropy concerning the redistribution process, i.e., the fast process, has been considered for a local stationary state of the process. In addition, it may be useful to note that the active noise with its temporal correlation in active matter systems in the equilibrium-like regime, which is treated as the fast variable, has acted as a source of nonequilibrium fluctuations in recent works in Refs. [14,15], where the entropy production is studied to see how the systems are far from equilibrium. Collisions of a Brownian particle with self-propelled particles in a thermal bath are modelled by such an active noise, for example.

We emphasize here that in our present discussion, the two variables mentioned earlier, i.e., $x$ and $y$, are of the slow dynamics.

For the entropic approach based on Eq. (1) to be applicable, the following consideration seems to be necessary. Regarding the above-mentioned regions as the imaginary blocks that can form a lot of collections of themselves, in each of which the conditional fluctuations are statistically equivalent to those in the original system, these blocks are independent each other in terms of $x$ (which will be mentioned again in Sections 4 and 5), so that a measure of uncertainty about the local property of the conditional fluctuations is introduced based on such collections: this measure is expected to become the conditional entropy in Eq. (1), in a usual manner similar to the one for deriving the Shannon entropy [2] (see Refs. [7-9] for relevant discussions).

In this article, we discuss a conditional entropic approach to nonequilibrium complex systems with a correlation between different fluctuations on a large time scale, which is weak in the sense that they are not completely statistically independent. Such a weak correlation is shown to constitute the conditional distribution $g(x|y_0)$, maximizing the entropy associated with the conditional fluctuations. We illustrate the approach in diffusion phenomena of histonelike nucleoid-structuring proteins (known as DNA-binding proteins) inside living *Escherichia coli* bacteria. In addition, we present its further possible illustration for p-granules, i.e., membraneless organelles in *C. elegans* embryos and beads in cell extracts derived from the eggs of *Xenopus laevis*, which share common statistical properties of fluctuations in their diffusion.

## 2. Weak correlation

Motivated by a recent work in Ref. [16] (see also Ref. [17]), suppose that $x$ and $y$ are not fully statistically independent, the notion of which is referred to as *weak correlation* between $x$ and $y$. To realize this, write the conditional distribution as $g(x|y) = e^{h(x|y)}$, where $h(x|y)$ is a suitable function and is approximately constant in terms of $y$ in the sense that it can be expanded around at $y = y_0$, with $y_0$ being the average value of $y$, up to the first order of $y - y_0$: $h(x|y) \cong h_0(x) + h_1(x)(y - y_0)$ with $h_0(x) \equiv h(x|y_0)$ and $h_1(x) \equiv h'(x|y_0)$, where the prime denotes differentiation with respect to $y$. $h_1(x)$ should be small in such a way that this expansion is valid in the whole range of $y$, meaning the linear approximation of $h(x|y)$ in terms of $y$. In other words, the degree of correlation in this case is small, i.e., weak, compared to that in the case when such an approximation is not legitimate. The former seems to be due to the nonequilibrium-stationary-state-like situation mentioned earlier, which is not in the strongly nonequilibrium regime. In order for the weak correlation to exist, it seems that the regions, each of which is in a local equilibrium state characterized by the dynamics of both $x$ and $y$, are weakly correlated each other, which may originate, for example, from the heat flux created by the temperature gradient between them, in general.

The weak correlation is also understood as follows. The change of $g(x|y)$ with respect to $y$, given a value of $x$, is here considered to be small, and correspondingly that of $h(x|y)$ is assumed to be very small since it is logarithmically related to $g(x|y)$, i.e., $h(x|y) = \ln g(x|y)$: the smallness is realized by $h_1(x)$, allowing us to perform the expansion only up to the first-order term. However, it may be necessary to take into account the higher-order terms, if $h(x|y)$ oscillates in terms of $y$, for example, although the expansion with the first-order term may hold for moderate oscillation. Now, one might think that the validity of such an expansion is not preserved in the case when $y$, which can approach infinity, takes large values far from $y_0$. A basic premise here is that $y_0$ is finite, which naturally implies that $f(y)$ tends to vanish at such large values [as can be seen in examples after Eq. (3) below], suggesting that the contribution from $g(x|y)$ becomes irrelevant since $g(x,y)$ under the expansion is expected to also have such a tendency due to its product structure. Therefore, in what follows, we treat a certain class of the conditional distributions, in which the validity of the expansion with the first-order term is preserved even for such a case.

Thus, the conditional distribution is found to be

$$g(x|y) \sim g(x|y_0)\exp[(y-y_0)h_1(x)]. \tag{2}$$

Clearly, $h_1(x)$ should not vanish for all the values of $x$ in its allowed range, since otherwise $\int dx\, g(x|y_0) e^{(y-y_0)h_1(x)} \sim 1$, which comes from the normalizability of $g(x|y)$, does not guarantee the normalization of $g(x|y_0)$.

Equation (2) offers the marginal distribution $g(x) = \int dy\, g(x,y)$ given by

$$g(x) \sim g(x|y_0)\exp[-y_0 h_1(x)]\int dy\, f(y)\exp[h_1(x)y]. \tag{3}$$

As shown in Refs. [16,17], the existence of the weak correlation is essential for describing the marginal distribution (see Section 4).

Closing this section, we present two examples to demonstrate the weak correlation. As the first example, let us consider the following bivariate exponential distribution [18]: $g(x,y) = e^{-x-y-\rho xy}[(1+\rho x)(1+\rho y)-\rho]$ defined on $[0,\infty) \times [0,\infty)$, where $\rho \in (0,1]$ is a small constant in the way mentioned above. This joint distribution leads to $g(x|y) = e^{-(1+\rho y)x}[(1+\rho y)(1+\rho x) - \rho]$ and $f(y) = e^{-y}$. Therefore, the weak correlation is given by $h_1(x) = \rho\{(1+\rho x)/[(1+\rho y_0)(1+\rho x) - \rho] - x\}$ with $y_0 = 1$ and vanishes only for $x = x^\dagger = [\sqrt{(1-\rho)^2 + 4\rho(1+\rho)} - (1-\rho)]/[2\rho(1+\rho)]$. Then, the second example is the following bivariate Gaussian distribution [19]:

$g(x, y) = [1/(2\pi\sqrt{1-\xi^2})]\exp\{-(x^2 - 2\xi xy + y^2)/[2(1-\xi^2)]\}$ defined on $(-\infty, \infty) \times (-\infty, \infty)$, where $\xi \in (-1, 1)$ is a nonzero small constant. In this case, the joint distribution yields $g(x|y) = [1/\sqrt{2\pi(1-\xi^2)}]\exp\{-(x-\xi y)^2/[2(1-\xi^2)]\}$ and $f(y) = (1/\sqrt{2\pi})\exp(-y^2/2)$, giving rise to $h_1(x) = [\xi/(1-\xi^2)](x - \xi y_0)$ with $y_0 = 0$, which vanishes only for $x = x^\dagger = 0$.

## 3. Maximization of conditional entropy and weak correlation

Clearly, the behavior of the conditional distribution on the left-hand side of Eq. (2) is dominantly determined by $g(x|y_0)$. In what follows, we show that $g(x|y_0)$ is realized in terms of the weak correlation $h_1(x)$, which is meant in the sense that the $x$ dependence of $h_0(x)$ comes only from $h_1(x)$ *itself* except possible additional quantities irrelevant to the weak correlation, if the maximization of the conditional entropy in Eq. (1) is considered. It is noticed that such a realization is not necessarily the case at the stage of Eq. (2), in general.

Under the assumption that $h_1(x)$ is given, the expansion of the exponential factor in Eq. (2) and the normalization condition on $g(x|y_0)$ lead to

$$\int dx g(x|y) \sim 1 + (y-y_0)\int dx g(x|y_0)h_1(x) + \sum_{n=2}^{\infty}[(y-y_0)^n/n!]\int dx\, g(x|y_0)[h_1(x)]^n,$$

where $[h_1(x)]^n$ is considered to be very small for $n \geq 2$, recalling the smallness of $h_1(x)$. Since the normalizability of $g(x|y)$ naturally requires the second term, which is dominant in terms of $h_1(x)$, on the right-hand side of this equation to vanish, we set the following condition:

$$\int dx\, g(x|y_0)h_1(x) = 0. \qquad (4)$$

Equation (4) means that the average of $h_1(x)$ should vanish and accordingly implies that $h_1(x)$ becomes positive or negative, depending on the values of $x$. Therefore, together with the constraint on the normalization condition, $\int dx g(x|y_0) = 1$, and a possible constraint on the average of a certain quantity, $Q(x)$, i.e., $\int dx\, g(x|y_0)Q(x) = \overline{Q}$, (and further constraints if any), the maximization of the conditional entropy in Eq. (1) reads

$$\delta_g\left\{S[g] - \lambda\left(\int dx g(x|y_0) - 1\right) + \nu\left(\int dx g(x|y_0)h_1(x) - 0\right)\right.$$

$$\left. + \kappa\left(\int dx g(x|y_0)Q(x) - \overline{Q}\right)\right\} = 0, \qquad (5)$$

where $\lambda$, $\nu$, and $\kappa$ are the set of the Lagrange multipliers concerning the constraints relevant to the normalization condition, the weak correlation, and the average value, respectively, and $\delta_g$ stands for the variation with respect to $g(x|y_0)$. The stationary solution of Eq. (5) is given by

$$\hat{g}(x|y_0) \propto \exp[\nu h_1(x) + \kappa Q(x)], \qquad (6)$$

which in fact shows that the conditional distribution *at* $y = y_0$ is realized by the weak correlation, highlighting its novel aspect. Later, the crucial importance of fluctuation distributions will be mentioned [see the discussion after Eq. (18) below].

A point to be emphasized here is that in contrast to the works [3-10] mentioned earlier, the quantity to be maximized is the conditional entropy associated with the slowly fluctuating variable, recalling the dominant role of $g(x|y_0)$ in Eq. (2).

**4. Protein diffusion in bacteria**

We here illustrate our approach of maximum conditional-entropy principle in the diffusion of histonelike nucleoid-structuring proteins inside living *Escherichia coli* bacteria observed in a recent experiment in Ref. [20] (see also Ref. [21]). Below, we will see that $x$ and $y$ correspond to describe the diffusion-exponent fluctuations and the temperature fluctuations, respectively.

The proteins distributed over the bacteria exhibit a highly heterogeneous diffusion in the sense that at the level of individual trajectories, the mean square displacement of a given protein scales for large elapsed time, $t$, as

$$\overline{\Delta x^2} \sim D_\alpha t^\alpha, \tag{7}$$

where $D_\alpha$ in units of $(\mu\text{m}^2\text{s}^{-\alpha})$ is the diffusion coefficient and its distribution obeys asymptotically the following power law

$$\varphi(D_\alpha) \sim D_\alpha^{-\gamma-1} \tag{8}$$

with $\gamma \cong 0.97$, whereas the diffusion exponent, $\alpha$, follows a non-trivial broad distribution in the range $0 \leq \alpha \leq 2$, see Figs. 2(d) and 2(c) in Ref. [20] (see also Figs. 1 and 2 in Ref. [16]), respectively. It is stressed here that Eq. (8) is the distribution for dimensionless numerical values of the diffusion coefficient, since its dimension changes, depending on the values of the diffusion exponent. Regarding the diffusion property, the case with $\alpha \neq 1$ $(\alpha > 0)$ is called anomalous diffusion [22], which is under vital investigation in the literature (see, e.g., Refs. [23-25] for recent reviews). It is noted that for small elapsed time, only normal diffusion, i.e., $\alpha = 1$, has been observed, for which the distribution of the diffusion coefficient, $D$, in units of $(\mu\text{m}^2\text{s}^{-1})$ decays as a power law, $\varphi(D) \sim D^{-\mu-1}$ with $\mu \cong 1.9$, see the left-bottom inset of Fig. 2(d) in Ref.

[20] (see also Fig. 3 in Ref. [16]).

In the experiment, the analysis of the mean square displacement in an ensemble average (i.e., an average of square displacement over all of the individual trajectories) has yielded both an average diffusion coefficient and an average diffusion exponent. What is remarkable there is the fact [20] that the former increases significantly, whereas the latter increases *only slightly*, with respect to the cell age (or, equivalently, cell length).

In Ref. [16], based on these experimental results and the assumption of the Einstein relation [26], i.e., $D \propto 1/\beta$ with the inverse temperature, $\beta \equiv 1/(k_B T)$, where $k_B$ is the Boltzmann constant, the concept of the weak correlation between the diffusion-exponent fluctuations and the temperature fluctuations has been introduced.

Therefore, in the present case, $x$ and $y$ correspond to $\alpha$ and $\beta$, respectively. The dynamics of both $\alpha$ and $\beta$ are much slower than the dynamics concerning stochastic motion of the protein in the local region, which is naturally taken as a typical local dynamics mentioned in the Introduction. Due to this slowness, the joint distribution of both $\alpha$ and $\beta$ is considered to be approximately constant on the time scale of such a local dynamics. With this hierarchical structure of time scales, $\alpha$ and $\beta$ slowly fluctuate locally over the bacteria. Moreover, according to the above-mentioned fact, $\alpha$-fluctuations may be slow compared to $\beta$-fluctuations since $\alpha$ seems to be approximately constant during the slow change of $\beta$. In this respect, it is of interest to experimentally examine this point.

We shall see the fluctuations in detail. In Ref. [16], in consistent with the power-law nature of the diffusion coefficient, the following inverse gamma distribution has been

proposed:

$$\varphi(D) \propto A^{\mu} D^{-\mu-1} \exp\left(-\frac{\mu A}{D}\right) \qquad (9)$$

in an interval, where $A$ is a positive constant with the dimension of the diffusion coefficient, see Fig. 3 in Ref. [16]. Under the assumption that the range of $\beta$ is unbounded, through the Einstein relation, this gives $\chi^2$ distribution for the temperature fluctuations [27] given by

$$f(\beta) = \frac{1}{\Gamma(\mu)} \left(\frac{\mu}{\beta_0}\right)^{\mu} \beta^{\mu-1} \exp\left(-\frac{\mu \beta}{\beta_0}\right), \qquad (10)$$

where $\beta_0$ is the average value of $\beta \in (0, \infty)$ and $\Gamma(\mu)$ is the Euler gamma function. Also, for the dimensionless numerical values of $D_\alpha$, the following inverse gamma distribution has also been suggested:

$$\varphi(D_\alpha) \propto \tilde{A}^{\gamma} D_\alpha^{-\gamma-1} \exp\left(-\frac{\gamma \tilde{A}}{D_\alpha}\right) \qquad (11)$$

in an interval, where $\tilde{A}$ is a dimensionless positive constant, in consistent again with the power-law nature, see Fig. 1 in Ref. [16]. As seen below, this distribution plays a crucial role for determining the conditional distribution. With the assumption, which seems to be experimentally supported [16] (see the discussion after Eq. (16) below and Ref. [28]), that $D_\alpha$ scales as

$$D_\alpha = D_{\alpha,\beta} \sim \frac{c}{s^\alpha \beta} \qquad (12)$$

with $s$ and $c$ being, respectively, a typical time characterizing the displacement of the protein and a positive constant, it has been suggested that given a value of $\beta$, the distribution of $D_\alpha$ takes also an inverse gamma distribution in Eq. (11) with $\tilde{A} = \tilde{A}(\beta)$. This gives the conditional distribution obtained by $g(\alpha|\beta) = |\partial D_{\alpha,\beta}/\partial \alpha| \varphi(D_\alpha)$ in the range $0 \leq \alpha \leq 2$, which behaves as

$$g(\alpha|\beta) = N(\beta) s^{\gamma\alpha} \exp\left[-\frac{\gamma a(\beta)}{c} s^\alpha\right], \qquad (13)$$

where $a(\beta)$ is a positive quantity depending weakly on $\beta$ provided that $\tilde{A}(\beta) = a(\beta)/\beta$ and $N(\beta)$ is the normalization factor calculated to be

$$N(\beta) = \frac{[\gamma a(\beta)/c]^\gamma \ln s}{\Gamma(\gamma, \gamma a(\beta)/c) - \Gamma(\gamma, s^2 \gamma a(\beta)/c)} \qquad (14)$$

with $\Gamma(k,r)$ being the incomplete gamma function defined by $\Gamma(k,r) = \int_r^\infty du\, u^{k-1} e^{-u}$.

Here, it is supposed that $a(\beta)$ can be expanded as $a(\beta) \cong a_0 + a_1(\beta - \beta_0)$ with $a_0 \equiv a(\beta_0)$ and $a_1 \equiv a'(\beta_0)$: $a_1$ is small to realize the weakness of correlation. In this case, $a_1$ is required to be negative in conformity with the cell-age dependence mentioned above [16]. Accordingly, $h_1(\alpha)$ takes the following form:

$$h_1(\alpha) = \frac{\gamma a_1}{c}\left(\langle s^\alpha \rangle_\alpha - s^\alpha \right), \qquad (15)$$

where $\langle \bullet \rangle_\alpha$ stands for the average with respect to $g(\alpha | \beta_0)$ over $\alpha \in [0, 2]$. It turns out that $h_1(\alpha)$ in Eq. (15) vanishes only when the value of $\alpha$ has $\alpha^\dagger = (\ln \langle s^\alpha \rangle_\alpha)/\ln s \cong 0.6$, where $s = 0.045\,\mathrm{s}$ [20]. This observation means that $h_1(\alpha)$ is positive or negative, depending on other values of $\alpha$, as pointed out in Section 3. With Eqs. (3), (10), (13), and (15), the marginal distribution $g(\alpha)$ is found to be given by

$$g(\alpha) \sim g(\alpha | \beta_0) \frac{\exp[-\beta_0 h_1(\alpha)]}{[1-(\beta_0/\mu)h_1(\alpha)]^\mu} \qquad (16)$$

with $1-(\beta_0/\mu)h_1(\alpha) > 0$ [16] in a good agreement with the experimental data, see Fig. 2 in Ref. [16]. This seems to support the validity of the expansion with the first-order term performed in Section 2. In Ref. [17], it has quantitatively been discussed how largely the conditional distribution is modulated by the weak correlation. Accordingly, it is of interest to examine if the conditional distribution with the weak correlation is observed in the experiments.

At this stage, regarding the relation in Eq. (12), we further mention the following. According to the experiments [20], at the level of the ensemble average mentioned above, the numerical value of the diffusion coefficient in the case of normal diffusion for small elapsed time, which is $24 \times 10^3 \, \text{nm}^2/\text{s}$, is three times larger than that in the case of anomalous diffusion for large elapsed time. As shown in Ref. [16], the average of $D_{\alpha,\beta}$ with respect to the distribution in Eq. (10) in the case of normal diffusion, i.e., $\alpha=1$, is calculated to be $\int_0^\infty d\beta \, f(\beta) D_{1,\beta} \sim [\mu/(\mu-1)][c/(s\beta_0)]$. Since this is relevant for large elapsed time, its value should approximately take one-third of the former value, from which the value of $\beta_0/c$ is estimated as $5.9 \times 10^{-3} \, \text{nm}^{-2}$ in a dimensionally consistent manner. The average of $D_{\alpha,\beta}$ with respect to the joint fluctuations is then given by $\int_0^2 d\alpha \int_0^\infty d\beta \, g(\alpha,\beta) D_{\alpha,\beta} \sim [\mu/(\mu-1)](c/\beta_0) \langle s^{-\alpha} [1-(\beta_0/\mu) h_1(\alpha)]^{1-\mu} \exp[-\beta_0 h_1(\alpha)] \rangle_\alpha$, the value of which is approximately found to be $10 \times 10^3 \, \text{nm}^2/\text{s}^\alpha$. Clearly, this value is close to $8.0 \times 10^3 \, \text{nm}^2/\text{s}^\alpha$, which is the average value of $D_\alpha$ measured in the experiments [20], and therefore this observation seems to support the assumption of the relation in Eq. (12).

Now, since the trajectories of the proteins are individual, a value of $\alpha$ is obtained from a given single trajectory, implying that the imaginary blocks mentioned in the Introduction are independent each other in terms of $\alpha$. Accordingly, our approach based on the conditional entropy in Eq. (1) seems to be applicable.

Therefore, taking the diffusion exponent $\alpha$ as the quantity $Q$, from Eqs. (6) and (15), we obtain the following conditional distribution:

$$\hat{g}(\alpha \,|\, \beta_0) \propto \exp[\nu\, h_1(\alpha) + \kappa\, \alpha]. \tag{17}$$

This becomes identical to $g(\alpha \,|\, \beta_0)$ in Eq. (13) after the following choices are made:

$$\nu = \frac{a_0}{a_1}, \qquad \kappa = \gamma \ln s, \tag{18}$$

where it is understood that all quantities are dimensionless. It is obvious that the condition in Eq. (4) is indeed fulfilled by Eq. (17) with Eq. (18).

Thus, we see that the weak correlation governs the conditional distribution in the present case of the protein diffusion in bacteria.

Closing this section, we wish to mention several works, in which the statistical property of fluctuations is of crucial importance for describing the displacement distribution. In Ref. [16], it has been shown, in view of superstatistics, that the temperature fluctuation distribution in Eq. (10) has successfully led to $q$-Gaussian distribution [29] (also called a Pearson-type VII distribution [30]), which decays as a power law, observed experimentally for displacements of the proteins [20] (see Fig. 4 in Ref. [16]): the fluctuation distribution is superstatistically incorporated into a process of fractional Brownian motion [31,32] that offers a unified description of anomalous diffusion as well as normal diffusion based on a fractional operator [33] (see Ref. [34] for a recent development about modelling anomalous diffusion toward such a direction). Thus, the conditional distribution in Eq. (13) should play a central role for describing the displacement distribution on a sufficiently long time scale, for which it is nonnegligible as discussed in Ref. [16]. The non-Gaussian diffusion has also been found to emerge from superstatistical frameworks with the distribution of the diffusion-coefficient/size fluctuations [35-38] or that of the diffusion-exponent fluctuations [39,40] (see also, e.g., Refs. [41-43]).

## 5. Diffusion of membraneless organelles in embryos and beads in cell extracts

Heterogeneous diffusion with fluctuations in the sense similar to Eq. (7) have also been observed in recent experiments in Refs. [44,45]. In this section, examining the gross behavior of these fluctuations, we present a further possible illustration of our approach.

Systems studied there are the following two different ones. One is of the p-granules in embryos of early *C. elegans* in Ref. [44], which are membraneless organelles and are known to form droplet-like assemblies of proteins and nucleic acids, and the other is the beads in cell extracts derived from the eggs of *Xenopus laevis* in Ref. [45], which are the cytosol of an ensemble of cells without large organelles. It has then been reported, from all of their individual trajectories examined there, that the diffusion-exponent fluctuation distribution takes a unimodal form, see Fig. 2(a) in Ref. [44] and Fig. 2(a) in Ref. [45], whereas the diffusion-coefficient fluctuation distribution obeys almost a lognormal type, see Fig. 2(b) in Ref. [44] and Fig. 2(b) in Ref. [45]. Regarding the latter, it is noted under the present symbols that given a value of $\alpha$, the values of $D_\alpha \times 1\mathrm{s}^\alpha$ (i.e., a typical spatial area within a period of $1\mathrm{s}$) have been evaluated in contrast to the dimensionless numerical values of the diffusion coefficient in Ref. [20].

Thus, these systems share common natures not only in the diffusion-coefficient fluctuations but also in the diffusion-exponent fluctuations. In what follows, assuming Eq. (12), we again consider that $x$ and $y$ correspond to $\alpha$ and $\beta$, respectively.

For the sake of simplicity, we use the same symbol $D_\alpha$ in Eq. (12) as the values of $D_\alpha \times 1\mathrm{s}^\alpha$, by which it is understood here and hereafter that $D_\alpha$ has the units of ($\mu\mathrm{m}^2$). From the experimental results in the above, the distribution of $D_\alpha$ we consider is the following lognormal distribution:

$$\varphi(D_\alpha) \propto \frac{1}{D_\alpha} \exp\left\{ -\frac{\left[\ln\left(D_\alpha / \tilde{A}\right)\right]^2}{2m^2} \right\} \qquad (19)$$

in an interval, where $\tilde{A}$ in this case and $m$ are positive constants. Like the discussions in Section 4, let us assume that given a value of $\beta$, the distribution of $D_\alpha$ has also a lognormal form in Eq. (19) with $\tilde{A} = \tilde{A}(\beta)$. Accordingly, from $g(\alpha | \beta) = \left| \partial D_{\alpha,\beta} / \partial \alpha \right| \varphi(D_\alpha)$ with $\tilde{A}(\beta) = a(\beta) / \beta$, in which the expansion mentioned before holds for $a(\beta)$ to realize the weak correlation, the conditional distribution in this case is found to behave as the following Gaussian form:

$$g(\alpha | \beta) = N(\beta) \exp\left\{ -\frac{\left[\alpha \ln s - \ln\left(c / a(\beta)\right)\right]^2}{2m^2} \right\} \qquad (20)$$

with $N(\beta)$ being the normalization factor given by

$$N(\beta) = \frac{(\ln s)/(m\sqrt{\pi/2})}{\mathrm{erf}\left(\left[\ln(c/a(\beta))\right]/(m\sqrt{2})\right) - \mathrm{erf}\left(\left[\ln(c/a(\beta))\right]/(m\sqrt{2}) - (\alpha_{\max} \ln s)/(m\sqrt{2})\right)} \quad (21)$$

in the range $0 < \alpha \leq \alpha_{\max}$, where $s$ takes a certain value less than unity and $\alpha_{\max}$ is the maximum value of $\alpha$ depending on the experimental data [44,45], and $\mathrm{erf}(r)$ is the error function defined by $\mathrm{erf}(r) = (2/\sqrt{\pi})\int_0^r du\, e^{-u^2}$. It is noted here that the condition, $c/a_0 < 1$, should hold in order for $g(\alpha|\beta_0)$ to have a peak.

Using Eq. (20), we have the following weak correlation:

$$h_1(\alpha) = \frac{a_1}{a_0}\left[\frac{\ln(c/a_0) - \alpha \ln s}{m^2} + \frac{g(0|\beta_0) - g(\alpha_{\max}|\beta_0)}{\ln s}\right]$$

$$= \frac{a_1}{m^2 a_0}\left[\ln\left(\frac{c}{a_0}\right) - \alpha \ln s\right], \quad (22)$$

where $g(0|\beta_0) = 0$ and $g(\alpha_{\max}|\beta_0) = 0$ have been used at the second equality, since they are reasonable from the experimental results of the scatter plot of the relation between $D_\alpha$ and $\alpha$, see Fig. 2(c) in Ref. [44] [where $D_\alpha$ in units of ($\mu\mathrm{m}^2\mathrm{s}^{-\alpha}$) has been used] and Fig. 2(c) in Ref. [45]. It is clear that the weak correlation vanishes only at $\alpha = \alpha^\dagger = [\ln(c/a_0)]/\ln s$, which should exist in the allowed range of $\alpha$.

In Refs. [44,45], no experimental results have been presented for the distribution of the diffusion coefficient in the case of normal diffusion. A point here is the experimental fact [44,45] that a typical value of $\alpha$ is close to unity, suggesting that the case of normal diffusion predominantly contributes to the realization of the distribution in Eq. (19). Accordingly, we suppose that the diffusion coefficient in the case of normal diffusion follows a lognormal distribution like in Eq. (19) and, through the Einstein relation, so does the distribution of $\beta$, i.e., $f(\beta)$. Substituting Eqs. (20), (22), and this distribution $f(\beta)$ into Eq. (3), we obtain the marginal distribution $g(\alpha)$, which is not analytically tractable. So, evaluating it as

$$g(\alpha) \sim g(\alpha \mid \beta_0) \left\{ 1 + \frac{1}{2} \sigma^2 \left[ h_1(\alpha) \right]^2 + \cdots \right\} \tag{23}$$

with $\sigma^2$ being the variance of $\beta$ with respect to $f(\beta)$, i.e., $\sigma^2 = \int d\beta \, f(\beta)(\beta - \beta_0)^2$, where the integration over $\beta$ has been taken after the expansion of the exponential factors in Eq. (3), we focus ourselves on a situation that the second term inside the square brackets on the right-hand side of Eq. (23) is small enough (due to the weak correlation) but still finite (due to the variance), which may be supported since $f(\beta)$ is expected not to be very sharply peaked. Therefore, we here treat the right-hand side of Eq. (23) up to the second order of $h_1(\alpha)$. Due to the unimodality of the diffusion-exponent fluctuation distribution [44,45], $g(\alpha)$ in Eq. (23) should have only one peak in the range of $\alpha$. This is indeed the case, as we shall see below.

The derivative of $g(\alpha)$ with respect to $\alpha$ leads, up to a proportionality constant and $g(\alpha \mid \beta_0)$, to $dg(\alpha)/d\alpha \sim [\alpha \ln s - \ln(c/a_0)] \{ \alpha^2 (\ln s)^2 / 2 - \alpha (\ln s) \ln(c/a_0) + [\ln(c/a_0)]^2 / 2 + (m^4 / \sigma^2)(a_0 / a_1)^2 - m^2 \}$, showing that its stationarity condition gives a real solution $\alpha = \alpha^{\dagger}$, at which $g(\alpha)$ is peaked, provided that $(m^2/\sigma^2)(a_0/a_1)^2 > 1$, which comes from the negativity of the discriminant of quadratic equation of $\alpha$ in the square brackets on the right-hand side: it is seen to be fulfilled since $[\ln(c/a_0)]^2/(2m^2)$ should be very large compared to unity due to $g(0|\beta_0) = 0$, the quantity on the left-hand side of this inequality is found to be larger than $1/\{(\sigma^2/2)[h_1(0)]^2\}$, the denominator of which is small enough, as mentioned above.

    Taking into account all the discussions in the above, let us apply our approach. Like in Section 4, due to the individual trajectories, again it seems that the approach is applicable since the imaginary blocks are expected to be independent of each other. Regarding the quantity $Q$, we shall impose a constraint on the size of the fluctuations of $\alpha^2$ (see Ref. [8] for a relevant discussion). From Eq. (6) with Eq. (22), therefore, we have the following conditional distribution:

$$\hat{g}(\alpha|\beta_0) \propto \exp[\nu h_1(\alpha) + \kappa \alpha^2]. \tag{24}$$

It is clear that this distribution becomes equivalent to $g(\alpha|\beta_0)$ in Eq. (20) after the following identifications are made:

$$\nu = \frac{a_0}{a_1}\ln\left(\frac{a_0}{c}\right), \qquad \kappa = -\frac{(\ln s)^2}{2m^2}, \tag{25}$$

with which the condition in Eq. (4) turns out to be satisfied.

Thus, in the present case, we again see that the conditional distribution is governed by the weak correlation.

## 6. Concluding remarks

We have developed a conditional entropic approach to nonequilibrium complex systems with a weak correlation between slowly fluctuating quantities. Maximizing the conditional entropy associated with such fluctuations, we have shown that the conditional fluctuation distribution is realized in terms of the weak correlation. We have also demonstrated the present approach for diffusion of histonelike nucleoid-structuring proteins in living *Escherichia coli* bacteria. Then, we have seen a further possible demonstration of the approach in diffusion of p-granules in embryos of early *C. elegans* as well as beads in cell extracts obtained from the eggs of *Xenopus laevis*.

Fluctuations in diffusion phenomena have experimentally been observed in other biological systems, e.g., as in Refs. [46-49]. The influence of protein conformational fluctuation to fluctuating diffusivity has been studied in Ref. [50]. In addition, a recent experimental work in Ref. [51] has reported a mild dependence of the diffusion exponent of telomeres in cells on temperature at the statistical level. This fact may imply the existence of the weak correlation between their fluctuations.

We make further some comments on relevant issues. The problems of diffusion concerning a growing granule or confinement have also been discussed in Refs. [52,53], respectively. It may be of interest to examine, in the present context, the deviation of the fluctuation distribution from its reference distribution [54] and a formal analogy of diffusivity fluctuations to thermodynamics [55], both of which are based on an entropic approach. A discussion has been made about (non-)Markovianity of the diffusion process (see, e.g., Ref. [56] for a relevant issue) that takes into account the diffusion-exponent fluctuations, for which the case of virus capsids in cells is treated [57]. In this respect, a recent experimental result about virus capsids in Ref. [58] is intriguing, where heterogeneous diffusion has been observed.

**Acknowledgments**

The author has been supported by a Grant-in-Aid for Scientific Research from the Japan Society for the Promotion of Science (No. 21K03394). This work has been completed while he has stayed at the Institut für Computerphysik, Universität Stuttgart. He would like to thank the Institut für Computerphysik for hospitality.

**References**

1. Khinchin, A.I. *Mathematical Foundations of Information Theory*; Dover: New York, USA, 1957.

2. *E.T. Jaynes: Papers on Probability, Statistics and Statistical Physics*; Rosenkrantz, R.D., Ed.; Kluwer: Dordrecht, Germany, 1989.

3. Crooks, G.E. Beyond Boltzmann-Gibbs statistics: Maximum entropy hyperensembles out of equilibrium. *Phys. Rev. E* **2007**, *75*, 041119.

4. Abe, S.; Beck, C.; Cohen, E.G.D. Superstatistics, thermodynamics, and fluctuations. *Phys. Rev. E* **2007**, *76*, 031102.

5. Van der Straeten, E.; Beck, C. Superstatistical distributions from a maximum entropy principle. *Phys. Rev. E* **2008**, *78*, 051101.

6. Hanel, R.; Thurner, S.; Gell-Mann, M. Generalized entropies and the transformation group of superstatistics. *Proc. Natl. Acad. Sci. USA* **2011**, *108*, 6390–6394.

7. Itto, Y. Heterogeneous anomalous diffusion of a virus in the cytoplasm of a living cell. *J. Biol. Phys.* **2012**, *38*, 673–679.

8. Itto, Y. Gaussian fluctuation of the diffusion exponent of virus capsid in a living cell nucleus. *Phys. Lett. A* **2018**, *382*, 1238–1241.

9. Itto, Y. Time evolution of entropy associated with diffusivity fluctuations: diffusing diffusivity approach. *Eur. Phys. J. B* **2019**, *92*, 164.

10. Davis, S. Conditional maximum entropy and superstatistics. *J. Phys. A: Math. Theor.* **2020**, *53*, 445006.

11. Beck, C.; Cohen, E.G.D. Superstatistics. *Physica A* **2003**, *322*, 267–275.

12. Wilson, A.G. *Entropy in urban and regional modelling*; Pion: London, UK, 1970.

13. Popkov, Y.S. Dynamic models of macrosystems with self-reproduction and their application to the analysis of regional systems. *Ann. Reg. Sci.* **1993**, *27*, 165–174.

14. Dabelow, L.; Bo, S.; Eichhorn, R. Irreversibility in Active Matter Systems: Fluctuation Theorem and Mutual Information. *Phys. Rev. X* **2019**, *9*, 021009.

15. Caprini, L.; Marconi, U.M.B.; Puglisi, A.; Vulpiani, A. The entropy production of Ornstein-Uhlenbeck active particles: a path integral method for correlations. *J. Stat. Mech.* **2019**, 053203.

16. Itto, Y.; Beck, C. Superstatistical modelling of protein diffusion dynamics in bacteria. *J. R. Soc. Interface* **2021**, *18*, 20200927.

17. Itto, Y.; Beck, C. Weak correlation between fluctuations in protein diffusion inside bacteria. *J. Phys.: Conf. Ser.* **2021**, *2090*, 012168.

18. Gumbel, E.J. Bivariate Exponential Distributions. *J. Am. Stat. Assoc.* **1960**, *55*, 698–707.

19. Kullback, S. *Information Theory and Statistics*; Dover: New York, USA, 1997.

20. Sadoon, A.A.; Wang, Y. Anomalous, non-Gaussian, viscoelastic, and age-dependent dynamics of histonelike nucleoid-structuring proteins in live *Escherichia coli*. *Phys. Rev. E* **2018**, *98*, 042411.

21. Sadoon, A.A.; Khadka, P.; Freeland, J.; Gundampati, R.K.; Manso, R.H.; Ruiz, M.; Krishnamurthi, V.R.; Thallapuranam, S.K.; Chen, J.; Wang, Y. Silver ions caused faster diffusive dynamics of histone-like nucleoid-structuring proteins in live bacteria. *Appl. Environ. Microbiol.* **2020**, *86*, e02479-19.

22. Bouchaud, J.-P.; Georges, A. Anomalous diffusion in disordered media: Statistical mechanisms, models and physical applications. *Phys. Rep.* **1990**, *195*, 127–293.

23. Höfling, F.; Franosch, T. Anomalous transport in the crowded world of biological cells. *Rep. Prog. Phys.* **2013**, *76*, 046602.

24. Metzler, R.; Jeon, J.-H.; Cherstvy, A.G.; Barkai, E. Anomalous diffusion models and their properties: non-stationarity, non-ergodicity, and ageing at the centenary of single particle tracking. *Phys. Chem. Chem. Phys.* **2014**, *16*, 24128–24164.

25. dos Santos, M.A.F. Analytic approaches of the anomalous diffusion: A review. *Chaos, Solitons and Fractals* **2019**, *124*, 86–96.

26. Nelson, P. *Biological Physics: Energy, Information, Life*; W.H. Freeman and Company: New York, USA, 2004.

27. Beck, C. Dynamical Foundations of Nonextensive Statistical Mechanics. *Phys. Rev. Lett.* **2001**, *87*, 180601.

28. Goychuk, I. Viscoelastic subdiffusion: From anomalous to normal. *Phys. Rev. E* **2009**, *80*, 046125.

29. Tsallis, C. *Introduction to Nonextensive Statistical Mechanics—Approaching a Complex World*; Springer: New York, USA, 2009.

30. Pearson, K. IX. Mathematical contributions to the theory of evolution.—XIX. Second supplement to a memoir on skew variation. *Phil. Trans. R. Soc. Lond. A* **1916**, *216*, 429–457.

31. Mandelbrot, B.B.; van Ness, J.W. Fractional Brownian motions, fractional noises and applications. *SIAM Rev.* **1968**, *10*, 422–437.

32. Goychuk, I.; Kharchenko, V.O. Rocking Subdiffusive Ratchets: Origin, Optimization and Efficiency. *Math. Model. Nat. Phenom.* **2013**, *8*, 144–158.

33. *Applications of fractional calculus in physics*; Hilfer, R., Ed.; World Scientific: Singapore, 2000.

34. Szarek, D.; Maraj-Zygmąt, K.; Sikora, G.; Krapf, D.; Wyłomańska, A. Statistical test for anomalous diffusion based on empirical anomaly measure for Gaussian processes. *Comput. Stat. Data Anal.* **2022**, *168*, 107401.

35. Chechkin, A.V.; Seno, F.; Metzler, R.; Sokolov, I.M. Brownian yet Non-Gaussian Diffusion: From Superstatistics to Subordination of Diffusing Diffusivities. *Phys. Rev. X* **2017**, *7*, 021002.

36. Maćkała, A.; Magdziarz, M. Statistical analysis of superstatistical fractional Brownian motion and applications. *Phys. Rev. E* **2019**, *99*, 012143.

37. Metzler, R. Superstatistics and non-Gaussian diffusion. *Eur. Phys. J. Spec. Top.* **2020**, *229*, 711–728.

38. Hidalgo-Soria, M.; Barkai, E. Hitchhiker model for Laplace diffusion processes. *Phys. Rev. E* **2020**, *102*, 012109.

39. Itto, Y. Heterogeneous anomalous diffusion in view of superstatistics. *Phys. Lett. A* **2014**, *378*, 3037–3040.

40. Itto, Y. Virus Infection Pathway in Living Cell: Anomalous Diffusion, Exponent Fluctuations, and Time-Scale Separation. In *Frontiers in Anti-Infective Drug Discovery*; Atta-ur-Rahman; Iqbal Choudhary, M., Eds.; Bentham Science

Publishers: Sharjah, United Arab Emirates, 2017; Volume 5.

41. dos Santos, M.A.F.; Junior, L.M. Random diffusivity models for scaled Brownian motion. *Chaos, Solitons and Fractals* **2021**, *144*, 110634.

42. dos Santos, M.A.F.; Colombo, E.H.; Anteneodo, C. Random diffusivity scenarios behind anomalous non-Gaussian diffusion. *Chaos, Solitons and Fractals* **2021**, *152*, 111422.

43. Korabel, N.; Han, D.; Taloni, A.; Pagnini, G.; Fedotov, S.; Allan, V.; Waigh, T.A. Unravelling Heterogeneous Transport of Endosomes. *arXiv* **2021**, arXiv:2107.07760.

44. Benelli, R.; Weiss, M. From sub- to superdiffusion: fractional Brownian motion of membraneless organelles in early *C. elegans* embryos. *New J. Phys.* **2021**, *23*, 063072.

45. Speckner, K.; Weiss, M. Single-Particle Tracking Reveals Anti-Persistent Subdiffusion in Cell Extracts. *Entropy* **2021**, *23*, 892.

46. Li, H.; Dou, S.-X.; Liu, Y.-R.; Li, W.; Xie, P.; Wang, W.-C.; Wang, P.-Y. Mapping Intracellular Diffusion Distribution Using Single Quantum Dot Tracking: Compartmentalized Diffusion Defined by Endoplasmic Reticulum. *J. Am. Chem. Soc.* **2015**, *137*, 436–444.

47. Stadler, L.; Weiss, M. Non-equilibrium forces drive the anomalous diffusion of telomeres in the nucleus of mammalian cells. *New J. Phys.* **2017**, *19*, 113048.

48. Etoc, F.; Balloul, E.; Vicario, C.; Normanno, D.; Liße, D.; Sittner, A.; Piehler, J.; Dahan, M.; Coppey, M. Non-specific interactions govern cytosolic diffusion of nanosized objects in mammalian cells. *Nature Mater.* **2018**, *17*, 740–746.

49. Sabri, A.; Xu, X.; Krapf, D.; Weiss, M. Elucidating the Origin of Heterogeneous Anomalous Diffusion in the Cytoplasm of Mammalian Cells. *Phys. Rev. Lett.* **2020**, *125*, 058101.

50. Yamamoto, E.; Akimoto, T.; Mitsutake, A.; Metzler, R. Universal Relation between Instantaneous Diffusivity and Radius of Gyration of Proteins in Aqueous Solution. *Phys. Rev. Lett.* **2021**, *126*, 128101.

51. Benelli, R.; Weiss, M. Probing local chromatin dynamics by tracking telomeres. *Biophys. J.* **2022**, *121*, 2684–2692.

52. Weber, P.; Bełdowski, P.; Bier, M.; Gadomski, A. Entropy Production Associated with Aggregation into Granules in a Subdiffusive Environment. *Entropy* **2018**, *20*, 651.

53. Gadomski, A. On (sub)mesoscopic scale peculiarities of diffusion driven growth

in an active matter confined space, and related (bio)material realizations. *BioSystems* **2019**, *176*, 56–58.

54. Itto, Y. Deviation of the statistical fluctuation in heterogeneous anomalous diffusion. *Physica A* **2016**, *462*, 522–526.

55. Itto, Y. Fluctuating Diffusivity of RNA-Protein Particles: Analogy with Thermodynamics. *Entropy* **2021**, *23*, 333.

56. Łuczka, J.; Hänggi, P.; Gadomski, A. Diffusion of clusters with randomly growing masses. *Phys. Rev. E* **1995**, *51*, 5762.

57. Itto, Y.; Bosse, J.B. Sojourn-time Distribution of Virus Capsid in Interchromatin Corrals of a Cell Nucleus. *Acta Phys. Pol. B* **2018**, *49*, 1941–1948.

58. Wu, Y.; Cao, S.; Alam, M.N.A.; Raabe, M.; Michel-Souzy, S.; Wang, Z.; Wagner, M.; Ermakova, A.; Cornelissen, J.J.L.M.; Weil, T. Fluorescent nanodiamonds encapsulated by *Cowpea Chlorotic Mottle Virus* (CCMV) proteins for intracellular 3D-trajectory analysis. *J. Mater. Chem. B* **2021**, *9*, 5621–5627.